\documentclass[a4paper, final, 10pt]{iopart}

\usepackage{amssymb}
\usepackage[cp1251]{inputenc}
\usepackage[english]{babel}
\usepackage{graphicx}
\usepackage{color}
\usepackage{bm}
\usepackage[breaklinks=true,colorlinks=true,linkcolor=blue,urlcolor=blue,citecolor=blue]{hyperref}

\newcommand{\ep}{\varepsilon}
\newcommand{\Lr}{\mathcal{L}}
\newcommand{\tDlt}{\tilde\Delta}

\newcommand{\rev}[1]{{\color{black}#1}}

\newcommand{\bmat}{\left(\begin{array}{cccc}}
\newcommand{\emat}{\end{array}\right)}
\newcommand{\eqref}[1]{(\ref{#1})}

\begin{document}

\title{Nonlocality and dynamic response of Majorana states in fermionic superfluids}



\author{I. M. Khaymovich $^{1,2}$,
        J. P. Pekola$^{3,4}$ and
        A. S. Melnikov$^{2,5}$}
  \address{$^1$ Max-Planck-Institut f{\"u}r Physik komplexer Systeme, N{\"o}thnitzer Stra{\ss}e 38, 01187 Dresden, Germany}
  \address{$^2$ Institute for Physics of Microstructures, Russian Academy of Sciences - 603950 Nizhny Novgorod, GSP-105, Russia}
  \address{$^3$ Low Temperature Laboratory, Department of Applied Physics, Aalto University School of Science, P.O. Box 13500, FI-00076 Aalto, Finland}
  \address{$^4$ Chair of Excellence of the Nanosciences Foundation, 23 rue des Martyrs, 38000 Grenoble, France}
  \address{$^5$ Lobachevsky State University of Nizhny Novgorod, 23 Prospekt Gagarina, 603950, Nizhny Novgorod, Russia}

\ead{ivan.khaymovich@pks.mpg.de}


\date{\today}

\begin{abstract}
We suggest a microscopic model describing the nonlocal ac response of a pair of Majorana states in fermionic superfluids
beyond the tunneling approximation.
The time-dependent perturbations of quasiparticle transport
are shown to excite finite period beating of the wavefunction between the distant Majorana states.
We propose an experimental test
to measure the characteristic time scales of quasiparticle transport through the pair of Majorana states defining, thus,
quantitative characteristics of nonlocality known to be a generic feature of Majorana particles.
\end{abstract}

\vspace{2pc}
\submitto{\NJP}
%
%

\maketitle

\section{Introduction}
Search for Majorana bound states (MBS) has recently become an active topic in the condensed matter community \cite{Review_Alicea2012, Review_Been2013, Review_Elliott2015}.
These exotic states are known to be characterized by the coinciding annihilation and creation operators.
This is why it is quite natural to look for such states in superconducting systems where
the order parameter $\Delta$ is known 
to mix particles (electrons) and anti-particles (holes) because of the Andreev scattering processes. 
Standard singlet superconductivity still does not allow the formation of this kind of excitations while the more exotic triplet
state can host MBS.
Among the available superfluids there exist only a few possible candidates for the triplet pairing such as He-3,
$Sr_2RuO_4$ and heavy fermion compounds \cite{volovik-silaev,kallin}.
Alternatively, the effective triplet pairing can be induced, e.g., in semiconducting nanowires
\cite{MBS_in_nanowires_Oreg2010, MBS_in_nanowires_Stanescu2011}
in the presence of rather strong spin-orbit coupling and external magnetic field.
Despite the clear and reliable observation of zero bias peaks (ZBP) in the differential conductance measurements \cite{ZBA_MBS_Mourik2012, ZBA_MBS_Das2012}
and on the change in the charge periodicity of conductance in Coulomb blockade regime \cite{Marcus2016}
consistent with the existence of MBS it would be extremely important
to probe other attributes of these states especially keeping in mind alternative explanations of the ZBP
based on Kondo physics \cite{ZBA_Kondo_Lee2012}.

  \begin{figure}[h]
  \center{
  \includegraphics[width=0.5\textwidth]{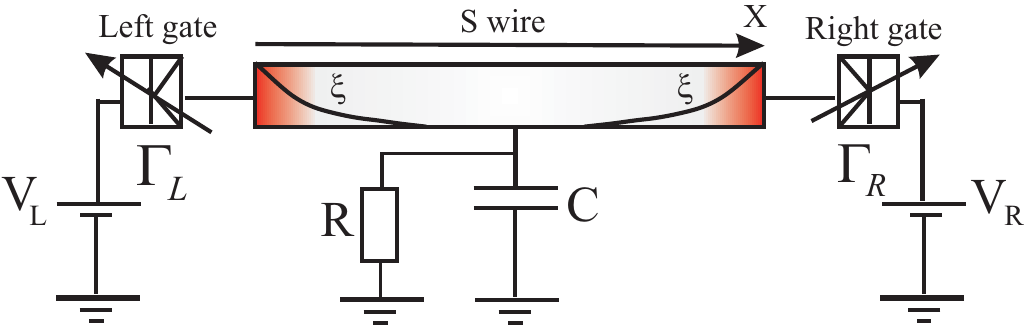}

  \caption{Setup of a possible experiment on
  Majorana dynamics.
  }
  \label{Fig:Setup}
  }
  \end{figure}

  The goal of this paper is to suggest a test revealing the nonlocal dynamic response of the MBS.
This issue has recently become a subject of intensive debate in the context of so-called quantum teleportation
\cite{MBS_teleport_Semenoff2007, MBS_teleport_Tewari2008, MBS_teleport_Fu2010, MBS_anti-teleport_BolechDemler2007, MBS_anti-teleport_NilsonAkhBeen2008}.
The Majorana partner states are localized at the length scales of the order of the coherence length $\xi$ and are usually strongly separated
provided the distance $L$ between them well exceeds this length $\xi$ (see Fig.~\ref{Fig:Setup}).
From the standard quantum mechanics one could naively expect that the time $\tau_0$ of the particle transfer between these localized states
should be determined by the inverse tunneling rate roughly proportional to
the value $\Delta e^{-L/\xi}$. Such scenario can be questioned if we remind that two Majorana states form a single fermionic level and, thus, the
injected particle should appear simultaneously in both partner states \cite{MBS_teleport_Semenoff2007, MBS_teleport_Tewari2008, MBS_teleport_Fu2010}.
This conclusion is in obvious contradiction with the analysis of the current noise correlations
\cite{MBS_anti-teleport_BolechDemler2007, MBS_anti-teleport_NilsonAkhBeen2008}: the latter points towards the existence of
a finite charge transfer time between the MBS.
Later on the teleportation phenomenon has been argued to be restored due to the nonlocal coupling via the Coulomb blockage \cite{MBS_teleport_Fu2010}.
It was concluded that the key omission of the previous studies was related to the
treating of the superconducting phase as a constant, and not as a dynamic variable. According to the work \cite{MBS_teleport_Fu2010}
the recovering of the nonlocal coupling between the MBS should occur if we consider
 the phase of the superconducting order parameter as a quantum variable canonically conjugate to the charge of the island.

In the present manuscript we show that the previous studies of the nonlocality in the system of the MBS suffer
from another key omission, namely they do not take into account the nonequilibrium effects responsible for the
mixing of the quasiparticle eigenfunctions with the positive and negative energies in the dynamic processes.
\rev{
}
In the remaining part of the paper we consider
a model describing the corresponding low frequency dynamics of the MBS
\rev{
}
and make clear predictions for the time-dependent experiment suggested above.
Specifically, our analysis demonstrates that the time of the quasiparticle transfer between  Majorana states
should be of the order of the inverse energy splitting $\tau_0\sim \omega_0^{-1}$ caused by their coupling $\omega_0$.
This result
\rev{
}
imposes restrictions on the time scales of adiabatic manipulation of the Majorana
states giving a criterion of their topological protection in time-dependent phenomena.
\rev{
For comparison it is interesting to mention here the work \cite{Dahan2017} where the dynamics is governed by time of flight of excitations in the normal metal wire coupled to the MBS.
}

\section{Model}
The low frequency dynamics of quasiparticles (QPs) can be described within the time-dependent
generalization of the BdG equations (cf. \cite{Ketterson_Song_book})
\begin{equation}\label{time-BdG}
i\frac{\partial}{\partial t}\hat g_n =
\bmat
\hat H_0-\mu& \hat\Delta\\ \hat\Delta^\dagger  & \mu-\hat H_0^*
\emat
\hat g_n \ .
\end{equation}
Here $\hat H_0$ is the normal state Hamiltonian, $\mu$ is the chemical potential, and $\hat g_n({\bf r},t)=(u_{\alpha,n},v_{\alpha,n})$.
The condition of adiabaticity naturally assumes that all the characteristic frequencies are much lower than the superconducting gap $\Delta$,
otherwise a full nonequilibrium description of a superconductor should be applied {\cite{Kopnin_book}}.
The coefficients $u_{\alpha,n}$ and $v_{\alpha,n}$ are usually interpreted as electronic- and hole- like parts of the
QP wave functions defined by the Bogolubov transformation,
\begin{eqnarray}
\label{psi}
\hat \Psi_{\alpha} ({\bf r},t) &=\sum\limits_n \left(u_{\alpha,n}({\bf r},t) \hat c_{n} +v^*_{\alpha,n}({\bf r},t) \hat c^\dagger _{n} \right) \ , \\
\hat \Psi^\dagger _{\alpha} ({\bf r},t) &=\sum\limits_n \left(u^*_{\alpha,n}({\bf r},t) \hat c^\dagger _{n} +v_{\alpha,n}({\bf r},t) \hat c_{n} \right) \ .
\end{eqnarray}
Here $\alpha$ is the spin index and $\hat c^\dagger _{n}$, $\hat c_{n}$ are the fermionic QP creation and annihilation operators, respectively.
The index $n$  enumerates the solutions of time-dependent BdG equations for different initial conditions
at $t=0$ when the expressions  (\ref{psi}) take the form of expansion
over a certain full set of functions.
In equilibrium the time dependence of the wave functions reduces to the standard form
$u_{\alpha,n}({\bf r},t)=\bar u_{\alpha,n}({\bf r})e^{-iE_n t}$, $v_{\alpha,n}({\bf r},t)=\bar v_{\alpha,n}({\bf r})e^{-iE_n t}$,
where $E_n$ and $(\bar u_{\alpha,n}({\bf r}),\bar v_{\alpha,n}({\bf r}))$ are the spectrum and eigenfunctions of the stationary BdG equations. Only the states with $E_n\geq 0$ contribute to the Eq.~(\ref{psi}) in this limit while in general the time-dependent solutions  $\hat g_n({\bf r},t)$ may contain the contributions from all positive and negative levels of the
stationary Hamiltonian.

The Majorana~-~type states in the stationary case can appear provided we have an isolated eigenfunction
satisfying the condition $v^*_{\alpha,0}=u_{\alpha,0}$ corresponding to zero energy.
The inverse transformation for this zero energy
state can specify only the sum of the fermionic operators
\begin{equation}
\frac{\hat c_{0}+\hat c^\dagger _0}{2} =\sum\limits_\alpha\int d{\bf r}\left(
 u^*_{\alpha,0}({\bf r}) \hat \Psi_{\alpha}({\bf r})  +u_{\alpha,0}({\bf r}) \hat \Psi^\dagger _{\alpha}({\bf r})
 \right)\ .
\end{equation}
This relation does not naturally yield the full fermionic operator $\hat c_0 = \hat \gamma_L+i\hat{\gamma}_R$ but
only its part $\hat \gamma_L=(\hat c_{0}+\hat c^\dagger _0)/2$ which indeed meets the Majorana conditions.
Another part ($\hat{\gamma}_R$) of the QP operator remains undefined and in this sense
the ground state of the superconductor with an isolated zero energy mode appears to be degenerate.
\rev{
}
The ambiguity of the operator $\hat{\gamma}_R$ can be resolved by introducing a coupling mechanism
of the above isolated state either to the second Majorana~-~type state or to a fermionic bath \cite{MBS_anti-teleport_NilsonAkhBeen2008, MBS_Buttiker_2012}.
Both these mechanisms destroy the symmetry of the isolated level $v^*_{\alpha,0}=u_{\alpha,0}$ and shift its energy from zero.
Each Majorana pair of states gives one positive and one negative energy level.
In equilibrium it is natural to keep only the positive energy level and the corresponding hybridized wave function. Considering the nonequilibrium dynamics at a finite time interval $t$ we can no more disregard the contribution of the negative energy level to the wave function dynamics when the energy uncertainty $\delta E\sim \hbar/t$ exceeds the splitting of levels in a Majorana pair. Thus, despite of
the obvious fact that both levels correspond to the only
fermion the nonequilibrium time-dependent solutions $\hat g_n({\bf r},t)$
of the BdG equations contain contributions corresponding to both levels.

\section{Nonequilibrium dynamics of a pair of Majorana states}
To probe the nonlocal dynamics of coupled Majorana states we suggest to study transport through the wire hosting these MBS at its ends modulated
by the changes in the coupling of the wire to the external normal metal leads (see Fig.~\ref{Fig:Setup}).
A natural way to tune this coupling in conditions of the real experiment (see, e.g., \cite{Hard_gap_Marcus2014, Epitaxial_growth_Marcus2015,Marcus2016})
is to apply time-dependent voltages at the gate electrodes controlling the transparencies $t_{L,R}(t)$ of the barriers between
the wire and the normal lead at the left (L) and right (R) end, respectively.
Tuning these transparencies at the two ends of the wire one can easily determine the spatial correlations in the dynamic response of the Majorana partners
as well as the scale $1/\tau_0$ of the frequency dispersion.
\rev{
Considering a possible experimental setup based on a semiconducting nanowire with induced superconductivity one should take this system in a topologically nontrivial state \cite{MBS_in_nanowires_Oreg2010, MBS_in_nanowires_Stanescu2011}
 which allows to get the subgap quasiparticle states bound to the wire ends.
}
Further derivation has been carried out
by applying a general approach \cite{Bardas_Averin_scatt_theory_many_harmonics_PRL_75_1831}
for the solution of the scattering problem with the quasiparticle waves incoming from the left or right leads
at a certain energy $\ep$
and propagating along the one-dimensional $p$-wave superconducting
wire hosting two MBS.
\rev{
We focus here on the case of a weak charging energy of the wire which is different from the situation studied in Ref.~\cite{MBS_teleport_Fu2010}.
}
The $p$-wave order parameter is chosen in the form $\Delta(x)\sim e^{i\theta_p}$, where $\theta_p=0,\pi$
 is the trajectory orientation angle.
Assuming low energies ($\ep,\omega_0\ll \Delta$)
and considering the solution of Eq.~\eqref{time-BdG}
near the left end of the wire one can write it
as a superposition
\begin{eqnarray}
g({\bf r},t) &= e^{-i\ep t+i k_F s}\left[a_L^+ w^{(1)}(s)+b_L^+ w^{(2)}(s)\right] \nonumber\\
&+e^{-i\ep t-i k_F s}\left[a_L^-w^{(1)}(-s)+b_L^-w^{(2)}(-s)\right] \
\end{eqnarray}
of two independent solutions
\begin{eqnarray}
w^{(1)}(s) =
e^{i\hat\sigma_z\theta_p/2}&\Bigl[e^{-D(s)/2}
\bmat
1\\ -i
\emat
\nonumber\\
&+i\frac{\ep}{\tilde \Delta}{\rm sign}(s)  e^{D(s)/2}
\bmat
1\\ i
\emat
\Bigr]
\ , \\
w^{(2)}(s) =
e^{i \hat\sigma_z\theta_p/2}&e^{D(s)/2}
\bmat
1\\ i
\emat
\ ,
\end{eqnarray}
found in Refs.~\cite{Vortex_tunneling_Kopnin_2003,Vortex_tunneling_Kopnin_2007} for the quasiclassical Andreev equations
at the trajectory with the coordinate $s = (L/2)\cos\theta_p + x$.
A similar expression can be written near the right end of the wire
by changing the subscripts $L\to R$ and the angle $\theta_p$ from $0$ to $\pi$, which shifts the origin $x\to x-L$ corresponding to $s(\theta_p=0)>0$ and $s(\theta_p=\pi)<0$.

Here $v_F$ is the Fermi velocity in the wire, $\tDlt^{-1} =\frac{2}{v_F} \int_0^{L/2} e^{-D(s)}ds$,
$D(s) =\frac{2}{v_F}\left|\int_0^{s}\Delta(s')ds'\right| \sim \frac{|s|}{\xi}$, and
Pauli matrices $\hat\sigma_k$ act in the electron~--~hole Gor'kov~--~Nambu space.
An appropriate matching of the wavefunctions at the wire ends
with the ones in the leads
gives us the equations for the coefficients $a_k^\pm = e^{\pm i \phi_k/2}(A_k \pm a_k)/2$ at the left ($k=L$) and right ($k=R$) wire ends (see \ref{App:Scattering} for details of calculations)
\begin{equation}\label{SM:A_k+a_k_eqs}
(\Gamma_k - i\ep)A_k = F_k \ ,\;
(\tilde\Delta - i\ep\Gamma_k/\tilde\Delta)a_k = F_k \ .
\end{equation}
Here for simplicity we neglect the MBS coupling $\omega_0\sim \tilde\Delta e^{-L/\xi}$,
$\Gamma_{k} = \tilde \Delta (1-r_k)/(1+r_k)$
is the rate characterizing the coupling of wire states to the $k$th external lead with
$r_k = \sqrt{1-|t_k|^2}$ being the real-valued reflection coefficient of the insulating barrier, $\phi_k$ are the scattering phases.
$F_{k}= \tilde \Delta t_k/(1+r_k)\propto \sqrt{\Gamma_k \tilde \Delta}$ are the tunneling sources characterizing the
incoming QP flows. Applying the Fourier transform with respect to the energy variable $\ep$
and considering the parameter $\omega_0/\tilde\Delta\sim e^{-L/\xi}$ pertubatively
one can obtain the equations describing the dynamics of a model two-level system in the time frame
 (cf.~\cite{Vavilov_Aleiner_scatt_time-depend_Green_func_2001,Quench_in_MBS_SIS_Been2015}), i.e.
 the dynamics of the Majorana pair:
 \begin{eqnarray}\label{Eq:time-depend-MBS-eqs_L}
\left( \frac{\partial}{\partial t} + \Gamma_L \right) A_L +\omega_0 A_R &= F_L e^{-i\ep t}\ ,
 \\
 \label{Eq:time-depend-MBS-eqs_R}
\left( \frac{\partial}{\partial t}+\Gamma_R \right) A_R -\omega_0 A_L &= F_R e^{-i\ep t}\ .
\end{eqnarray}
In the non-stationary regime the localized states at the wire ends (being of Majorana nature in the stationary regime) can be described  by the wave function amplitudes $A_k$ which are in fact the quantum mechanical amplitudes describing the probability to find the quasiparticle at the $k$th wire end.
The amplitudes $a_k$ correspond to the off-resonant fast-decaying contributions from the states above the gap.
The amplitudes $A_k$ and $a_k$ together describe in fact the low frequency dynamics of the function
$\hat g_n({\bf r},t)$ including contributions from positive and negative levels of the
stationary Hamiltonian.
Note that in the absence of incoming QP flows, $F_k=0$, Eqs.~(\ref{Eq:time-depend-MBS-eqs_L}, \ref{Eq:time-depend-MBS-eqs_R}) have purely real-valued coefficients corresponding to the Hermitian nature of  Majorana operators $\hat \gamma_k$.
In this case the average $\langle \Psi_\alpha^\dagger({\bf r},t) \Psi_\alpha ({\bf r},t)\rangle$ of the electron number operator is conserved since its change is determined by the sum $|A_L|^2 + |A_R|^2$ of probabilities $|A_k|^2$ to find the quasiparticle at the $k$th wire end. This conservation fixes, in particular, the quasiparticle parity number in the wire by fixing the parameter $|A_L|^2 + |A_R|^2$ even for non-trivial dynamics of $|A_k|^2$ themselves.
Note that this statement is independent of a strength of Coulomb interaction as the latter only governs the correlations between tunneling rates.
The rates $\Gamma_{L,R}$ are determined by the local Andreev reflection processes { \cite{MBS_anti-teleport_NilsonAkhBeen2008}}
while the energy splitting of coupled Majorana states  $\omega_0 = \tilde\Delta e^{-D(L/2)} \sin \varphi$
is related to the probability of the quasiparticle transfer through the system.
Parameters $D(L/2)\sim L/\xi$ and $\varphi = k_F L+(\phi_L-\phi_R)/2$
depend on the wire length $L$.

The current flowing from the left and right electrodes can be calculated as \cite{BTK} (see also \ref{App:dI_dV} for details of calculations)
\begin{equation}
I_{L,R}=e/\pi\int g_{L,R}(\ep)(f_T(\ep-eV_{L,R})-f_T(\ep-eV_s))d\ep \ ,
\end{equation}
where $f_T(\ep) = (e^{\ep/T}+1)^{-1}$ is the Fermi-Dirac distribution function with the bath temperature $T$,
\begin{equation}
g_{k}(\ep)\simeq 2{\rm Re}[A_{k} a_{k}^*]=2\sqrt{\Gamma_{k}}Re(A_{k}e^{i\ep t}) \ ,
\end{equation}
$V_{k}$ is the potential of the $k$th electrode,
and $V_s$ is the potential of a superconductor. Generally, the definition of the potential $V_s$ in a nonstationary
problem follows from the solution of the equations describing the particular electric circuit \cite{Hekking_Buttiker_floating_SC}, e.g., the one in Fig.~\ref{Fig:Setup}:
$I_L+I_R=C dV_s/dt+V_s/R$, where $C$ and $R$ are the capacitance and shunt resistance of the ground connection, respectively.
Considering a constant applied bias $V=V_L-V_R$ and putting $A_{L,R}\propto e^{-i\ep t}$ we obtain a dc differential conductance peak at $e V \simeq \omega_0$
attributed to MBS \cite{ZBA_MBS_Mourik2012, ZBA_MBS_Das2012,egger,book,Ioselevich_Feigelman2013}.

\section{Results}
We now proceed with the analysis of the dynamic response of a pair of Majorana partners and
 consider two generic examples of the time -- dependent transport realized by
  the modulating tunnel barrier (see Fig.~\ref{Fig:Setup}):
(i) the phase-shifted sinusoidal driving with $\Gamma_L(t) = \Gamma_0 +\tilde\Gamma\cos(\omega t)$ and
$\Gamma_R(t) = \Gamma_0 +\tilde\Gamma\cos(\omega t+\phi_0)$;
 (ii) pump-probe driving
by $\Delta t$-broadened delta-functional pulses with different amplitudes $G_k^t$ applied with a time delay $\tau$, i.e., with
$\Gamma_k(t) = G_k^0 \delta_{\Delta t}(t)+G_k^\tau \delta_{\Delta t}(t-\tau)$.

To start with, our consideration of the dynamic response of MBS within Eqs.~(\ref{Eq:time-depend-MBS-eqs_L}, \ref{Eq:time-depend-MBS-eqs_R})
\rev{
}
through a single fermionic state formed of a superposition of two partner Majorana states. Indeed, the levels
$\pm\omega_0$ around the zero energy
can be introduced as a basis of hybridized states with the amplitudes  $A_\pm=A_L\pm i A_R$.
In Eqs.~(\ref{Eq:time-depend-MBS-eqs_L}, \ref{Eq:time-depend-MBS-eqs_R})
each of quasiparticle sources $F_k$ excites both amplitudes $A_\pm$ simultaneously.
Due to the coupling to the reservoirs
both amplitudes evolve then in time as separate quantities and, thus, cannot be described
as an empty and filled state of a single level.
As a result,
\rev{
}
we find beating of the wavefunction
between the edge states at the frequency $\omega_0$.
The above arguments concerning the sources of the injected particles
should be valid irrespective to the
strength of the Coulomb effects and for the sake of simplicity we start our consideration
of time~-~dependent problems from the limit of large capacitance $C$ when these effects can be neglected.

Starting from the case of sinusoidal driving we consider for simplicity the ac amplitude $\tilde\Gamma\ll\Gamma_0$ as a perturbation and solve Eqs.~(\ref{Eq:time-depend-MBS-eqs_L}, \ref{Eq:time-depend-MBS-eqs_R}).
For the zero-bias differential conductance we find
\begin{eqnarray}\label{Eq:dIdV-sin-drive}
\left.\frac{1}{G_{\Gamma}}\frac{dI_L}{dV_L}\right|_{V_L=0}\simeq 1+\frac{\tilde \Gamma}{2\Gamma_0}
\Bigl[\cos\omega t\nonumber\\
+\frac{\omega_0^2-\Gamma_0^2}{2\Gamma_0}\sum_{\eta=\pm 1}\Lr_{\eta} F^\eta_0
-{\omega_0}\sum_{\eta=\pm 1}\eta\Lr_{\eta} F^\eta_{\phi_0-\pi/2}
\Bigr] \ ,
\end{eqnarray}
where $G_{\Gamma} = (e^2/\pi)2\Gamma_0^2/(\Gamma_0^2+\omega_0^2)$, $\Lr_\eta = \Gamma_0/[(\omega+\eta\omega_0)^2+\Gamma_0^2]$, $F^\pm_\phi = \cos (\omega t + \phi) + \sin (\omega t +\phi) (\omega\pm\omega_0)/\Gamma_0 $.
One can see that for low-frequencies $\omega\lesssim \omega_0$ the above expression contains an essential phase $\phi_0$ dependence,
while with increasing $\omega$ these contributions decay faster than the other time-dependent terms.

 \begin{figure*}[t]\center{
  \includegraphics[width=0.8\textwidth]{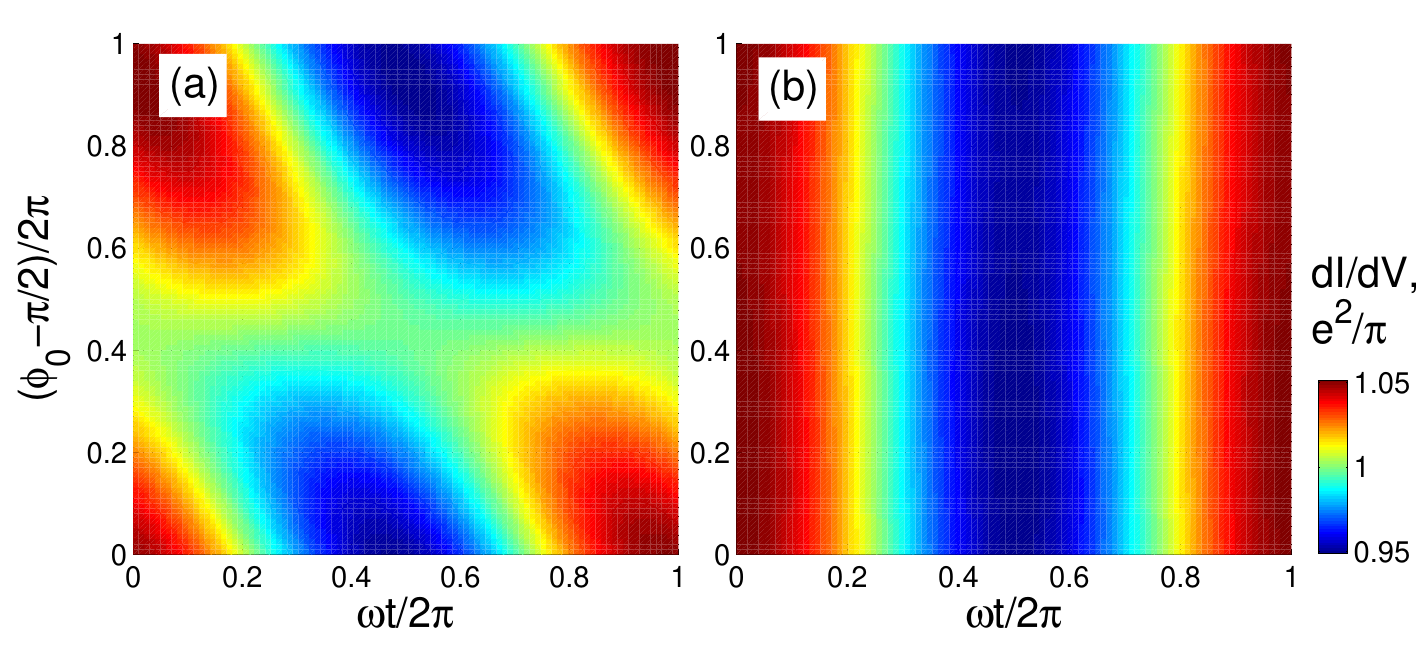}
  \caption{ Color plot of differential conductance \eqref{dI_dV_res} versus time $t$ and phase difference $\phi_0$ for (a) $\omega=\omega_0$ and (b) $\omega = 10\omega_0$.
  The other parameters are $\Gamma_0 = \omega_0 = 10\tilde\Gamma$. One can see strong phase dependence at $\omega\sim\omega_0$, which diminishes as $\omega$ grows.}
  \label{SM-Fig:dI_dV_sin}
  }
 \end{figure*}

Indeed, this statement is clearly visible in the most interesting and representative case $\omega_0\sim \Gamma_0$ in which dc results \cite{egger,book,Ioselevich_Feigelman2013} (see also \eqref{dc_dI_dV} in \ref{App:dc_dI_dV}) are already broadened and inconclusive.
In this case to clarify the results we rearrange the functions $F^\pm_\phi = \cos (\omega t + \phi) + \sin (\omega t +\phi) (\omega\pm\omega_0)/\Gamma_0  = F^c_\phi \pm F^s_\phi$ to
$F^c_\phi = \cos(\omega t +\phi) + (\omega/\Gamma_0)\sin (\omega t +\phi)$ and
$F^s_\phi = (\omega_0/\Gamma_0)\sin (\omega t +\phi)$
getting
\begin{eqnarray}\label{dI_dV_res}
\left.\frac{1}{G_{\Gamma}}\frac{dI_L}{dV_L}\right|&_{V_L=0}\simeq
1+\frac{\tilde \Gamma}{2\Gamma_0}\cos\omega t\nonumber\\
&+\frac{\tilde \Gamma}{\Gamma_0}
\Bigl[\frac{\omega_0^2-\Gamma_0^2}{2\omega_0^2}\left\{C_0 F^c_0-C_1 F^s_0\right\}\nonumber\\
&-\frac{\Gamma_0}{\omega_0}\left\{C_0 F^s_{\phi_0-\pi/2}-C_1 F^c_{\phi_0-\pi/2}\right\}
\Bigr]
\ ,
\end{eqnarray}
with the dimensionless coefficients $C_{k} = ({\pi \omega_0^2}/{2\Gamma_0}) \sum_{\eta=\pm 1}(-\eta)^k\Lr_\eta$
accumulating the $\omega$-dependence of the prefactors
as follows
\begin{eqnarray}
C_0(\omega) = \frac{(\omega^2+\omega_0^2+\Gamma_0^2)\omega_0^2}{[(\omega+\omega_0)^2+\Gamma_0^2][(\omega-\omega_0)^2+\Gamma_0^2]}, \\
C_1(\omega) = \frac{2\omega\omega_0^3}{[(\omega+\omega_0)^2+\Gamma_0^2][(\omega-\omega_0)^2+\Gamma_0^2]} \ .
\end{eqnarray}

Now considering two limits:
(a) $\omega\sim\omega_0,\Gamma_0$ and (b) $\omega\gg\omega_0,\Gamma_0$ illustrated in the corresponding panels of Fig.~\ref{SM-Fig:dI_dV_sin},
one can see that in the first limit (a) $\omega\sim\omega_0,\Gamma_0$ 
the above mentioned coefficients $C_{0,1}\sim 1$ weakly depend on the frequency $\omega$
and the phase dependent corrections (the last line in \eqref{dI_dV_res}) are of order of the main term $\frac{\tilde \Gamma}{2\Gamma_0}\cos\omega t$.

In the second limit (b) due to the smallness of $C_0 \simeq \omega_0^2/\omega^2$ and $C_1\simeq 2\omega_0^3/\omega^3$
the conductance has relatively small $\omega_0^2/\omega^2$ phase-dependent corrections to the oscillating terms
\begin{eqnarray}
\left.\frac{1}{G_{\Gamma}}\frac{dI_L}{dV_L}\right|_{V_L=0}\simeq
\frac{\tilde \Gamma}{2\Gamma_0} &\Bigl[\cos\omega t
+\frac{\omega_0^2-\Gamma_0^2}{\omega^2}F^c_0\nonumber\\
&-\frac{2\omega_0\Gamma_0}{\omega^2}F^s_{\phi_0-\pi/2}
\Bigr]
\ .
\end{eqnarray}
In Fig.~\ref{SM-Fig:dI_dV_sin} the $(t,\phi)$-dependence of the differential conductance \eqref{dI_dV_res} is plotted for the following parameters $\omega_0=\Gamma_0 = 10\tilde \Gamma$ at (a) $\omega = \omega_0$ and (b) $\omega=10\omega_0$ demonstrating the above mentioned arguments.

In the other limit of $\omega_0\gg \Gamma_0$
many beating periods pass before a tunneling event occurs leading to the efficient transport of the charge between the localized states $A_k$.
This can be in some sense viewed as a signature of ``teleportation''.
If additionally
$\Delta\omega = \omega-\omega_0\sim \Gamma_0\ll \omega_0$
one can neglect the contributions from $\eta=+1$ and obtain
\begin{eqnarray}
\left.\frac{\pi}{e^2}\frac{dI_L}{dV_L}\right|_{V_L=0}\simeq \frac{2\Gamma_0^2}{\omega_0^2}+\frac{\tilde \Gamma\Gamma_0}{\omega_0^2}\cos\omega t\nonumber\\
+\frac{\tilde \Gamma}{2\Gamma_0}\frac{\Gamma_0^2}{\Delta\omega^2+\Gamma_0^2}\Bigl[
F^-_0
+\frac{2\Gamma_0}{\omega_0}F^-_{\phi_0-\pi/2}
\Bigr] \ .
\end{eqnarray}
Clearly this limit describes the sharp peaks at $\omega_0$ in the frequency dependence
of the dynamic response with the amplitude that depends on the phase shift.
In the opposite limit of broad peaks the nonlocal correlations in the dynamic response
are naturally more difficult to observe since their contributions
in the dynamic response become small when $\omega_0/\Gamma_0\ll 1$.

For arbitrary bias and drive amplitudes we should get a multiplication of harmonics and considering the current averaged over the drive period we can expect the appearance of the conductance peaks at voltages $eV_{L,R}= n\omega\pm\omega_0$ due to the resonant effect similar to the Shapiro phenomenon in Josephson junctions { \cite{Tinkham_book}}.
Note that the periodic backgate voltage modulation can give
another opportunity to observe the resonant features
on the current -- voltage curve controlling the chemical
potential of the wire as a whole.
This modulation should cause the change in the energy splitting $\omega_0$ through its dependence on the Fermi momentum $k_F$.
Assuming $k_F L = \omega t$ to be linear in time one can obtain resonances at $e V_{L,R} = n \omega$.

\begin{figure}[h]
  \center{
  \includegraphics[width=0.45\textwidth]{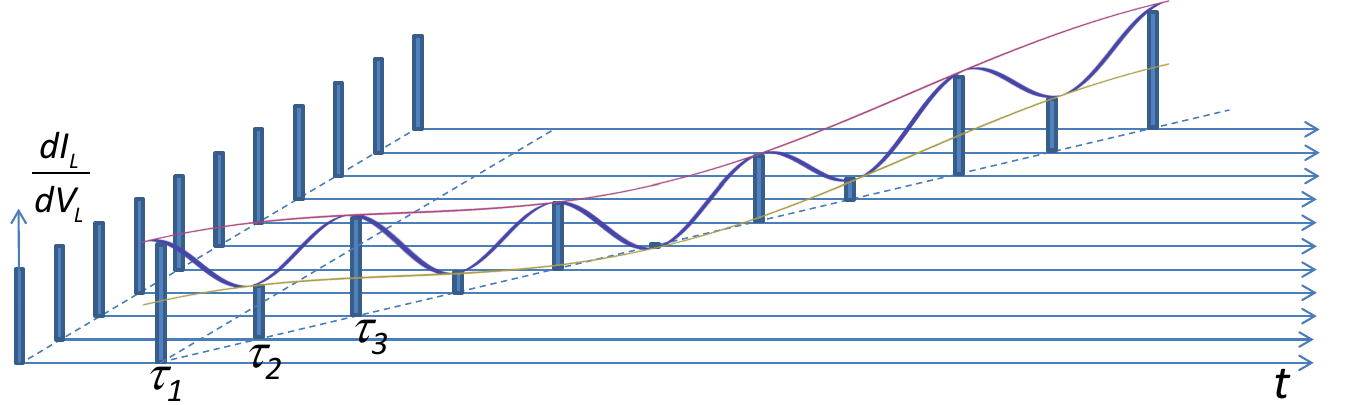}

  \caption{  Differential conductance vs delay time $\tau$ in two-pulse pump-probe
setup.
The second pulse amplitude is shown by solid blue line.
}
  \label{Fig:Pump-probe}
  }
  \end{figure}

In the case of the pump-probe driving the differential conductance of the left electrode contains three contributions
\begin{eqnarray}\label{g(E)-pump-probe}
\frac{\pi}{e^2 \Delta t}\frac{dI_L}{dV_L} = G_L^0 \delta_{\Delta t}(t)+G_L^\tau\delta_{\Delta t}(t-\tau)+\nonumber\\
\sqrt{G_L^0 G_L^\tau} \cos (\omega_0 \tau)\cos(e V_L \tau) \delta_{\Delta t}(t-\tau)
 \ .
\end{eqnarray}
Here we impose zero initial conditions on both amplitudes $A_\pm$.
The terms in the first line of Eq.~\eqref{g(E)-pump-probe} correspond to the local charging of the single fermionic level,
while the term on the second line reflects correlations in the response to two pulses with the time delay $\tau$
and shows the non-trivial dynamics of MBS at frequencies $|eV_L \pm \omega_0|$ (see Fig.~\ref{Fig:Pump-probe}).
The first pulse excites the quantum beatings between the Majorana edge states at the frequency $\omega_0$ modifying
the response of the system to the second pulse.

Taking for the estimate $\Delta \sim 2.5$~K, $\xi\sim 100$~nm for Al, and $L\gtrsim 1$~$\mu$m we find $\omega_0\lesssim 15$~MHz which gives us
a reasonable range of frequencies $\omega\sim \omega_0$ of the drive and typical time delay $1/\omega_0\sim 0.06$~$\mu$s for the pump-probe setup.
The conditions on bias for the observation of the beating phenomenon are less restrictive comparing to the ones of a dc conductance peak (with the restriction of $V\sim \omega_0\sim 0.01$~$\mu$V \cite{Ioselevich_Feigelman2013}) as ac measurements are already conclusive at zero bias, see, e.g., Eq.~\eqref{Eq:dIdV-sin-drive}.
To get $\Gamma_{L,R}\lesssim \omega_0$ we should take the barriers with resistances $R_{L,R}\sim 0.1-1 $~G$\Omega$.

\section{Discussion and outlook}
Certainly, the above dynamic response of the MBS will be modified in Coulomb blockade regime.
 This difference arises from the obvious fact that in the case of Coulomb blockade the charge tunneling processes between the island and left/right electrodes are strongly correlated. The entry and exit of charged particles are always controlled by the overall charge of the island. However, this correlation doesn't destroy the beating phenomenon and cannot cause the formation of a single eigenstate responsible for the non-local transport through Majorana states (teleportation) for the operating frequencies above the energy splitting of Majorana partners.
Let us take the limit of high Coulomb energy and imposing, thus, the restriction on two possible charge states of the island and assume the operating frequencies and energy splitting $\omega_0$ to be small comparing to the tunneling rates $\Gamma_{L,R}$. The latter limit allows one to consider the charging/discharging processes as instantaneous events changing the fermion parity.
On the longer time scales than the injection/ejection rates the fermion parity is fixed due to the fixed electron charge.
However, the beating phenomenon as an internal dynamics of Majorana states is present due to the nonequilibrium time-dependent nature of the electron injection and further transformation of the wave function of the injected electron into the Andreev eigenstates both with positive and negative energies. Therefore the
current through the system is fully determined by the interplay of two time scales, namely, the inverse beating frequency $\omega_0^{-1}$ and the delay time $\tau$ between the opening of the left/right junctions. The latter is determined either by the operating frequency $f$ and the phase shift $\phi$ ($\tau=(n+\phi/2\pi)/f$ with an integer $n$ value) for the periodic driving or by the delay time $\tau$ for the pump-probe experimental setup.
\rev{
Certainly the above comment on the influence of Coulomb blockade on the beating phenomenon is only qualitative and
should be verified by further quantitative analysis based on the use of more elaborated methods taking account of the interaction effects.
}

To conclude
\rev{
the solution of the above dynamic problems allows us to predict a beating effect
at the frequency $\omega_0$ which is a hallmark of the topologically nontrivial state of the nanowire.
We show that due to the exponentially small coupling $\omega_0$ the MBS are strongly sensitive
to any external perturbation.
}
According to our consideration any driving of Majorana states with the typical operating frequency $\omega$
exceeding $\omega_0$ brings the system to the non-equilibrium regime imposing, thus,
an important restriction on the operating frequencies of such a device
\rev{
}
The Majorana nature of these states needed for quantum calculations recovers only in the adiabatic regime $\omega\ll\omega_0$.
On the other hand, the
measurement of the characteristic frequency
threshold $\omega_0$ separating the regimes of weak and strong perturbations of the Majorana pairs could be
considered as their hallmark characterizing the nonlocality of these pairs.
\rev{
}
Certainly the beating phenomenon similar to the one
discussed in our work should appear in other superconducting systems with subgap Andreev
states. To distinguish the beating phenomenon in topological situation from
the one caused by the presence of usual Andreev states it may be
helpful to study the
behavior of the beating frequency as a function of system parameters,
gate potentials and magnetic field so that to reveal the features peculiar to
the topologically protected levels. 
The beating phenomenon may also affect non-stationary Josephson-type transport in systems with MBS studied in recent experiments \cite{Molenkamp_arx2016,Molenkamp_arx2017}.

\ack
We are pleased to thank A. A. Bespalov, Yu. G. Makhlin, C. Marcus, G. E. Volovik, and A. D. Zaikin for valuable comments
and A. J. Leggett for correspondence.
This work has been supported in part
by Microsoft Project Q,
by the Nanosciences Foundation, foundation under the aegis of the Joseph Fourier University Foundation (Grenoble, France),
by Academy of Finland 
Projects No. 284594, 
272218 (J.~P.~P.), 
by the Russian Foundation for Basic Research and German Research Foundation (DFG) Grant No. KH 425/1-1 (I.~M.~K.), and
by the Russian Science Foundation, Grant No. 17-12-01383 (A.~S.~M.)
\rev{
and Foundation for the advancement of theoretical physics “BASIS”.
}

\appendix
\section{Derivation of Eqs.~(\ref{Eq:time-depend-MBS-eqs_L}, \ref{Eq:time-depend-MBS-eqs_R})}\label{App:Scattering}
In this section we present the derivation of the  Eqs.~(\ref{Eq:time-depend-MBS-eqs_L}, \ref{Eq:time-depend-MBS-eqs_R}) from the main text for an exemplary system consisting of
 a one dimensional (1D) $p$-wave superconducting (S) wire of the length $L$ connected to the left and right one-dimensional normal-metal 
 leads. We choose the $x$ axis along the wire, the origin to be in the middle of the wire and the order parameter in the form
 $\Delta\propto \hat k_x + i \hat k_y$. Such system is known to host the subgap edge states at rather small energies
 $\pm\omega_0\sim \pm\Delta e^{-L/\xi}$.
To describe these localized states we start from the quasiclassical version of the Bogolubov~--~de~Gennes equations,
i.e., Andreev equations for the envelopes $w = (u, v)$ of the electron and hole waves propagating
 along the quasiclassical trajectory ${\bf k} = k_F (\cos\theta_p,\sin\theta_p)$
\begin{equation}\label{SM:Andreev_eqs}
-i v_F \hat \sigma_z\frac{\partial}{\partial s} w + \hat\sigma_x \Delta(s) w = \ep w
\end{equation}
where $v_F$ is the Fermi velocity in the wire, $s = (L/2)\cos\theta_p + x$ is the coordinate in the wire along the trajectory, and
Pauli $\hat\sigma_k$ matrices act in the electron~--~hole Gor'kov~--~Nambu space,
Considering the $p$-wave symmetry of superconducting order parameter one can put $\Delta(x)\sim e^{i\theta_p}$.
Note that in 1D geometry of the $p$-wave S wire it is natural to align the trajectory in the positive or negative direction of the $x$ axis
which correspond to $\theta_p=0,\pi$.
The phase $\theta_p$ can be removed from the gap operator $\Delta$ by the standard transformation $u(x)\to u(x)e^{i\theta_p/2}$ and $v(x)\to v(x)e^{-i\theta_p/2}$.

\subsection{Low energy modes inside the wire}
Considering the low energy modes with $\ep\ll \tDlt \sim \Delta$ inside the wire
one can
take the sum of two independent solutions $w^{(1,2)}(s)$ of Andreev equations \eqref{SM:Andreev_eqs}
found in Refs.~\cite{Vortex_tunneling_Kopnin_2003,Vortex_tunneling_Kopnin_2007}
\begin{eqnarray}
w^{(1)}(s) =
e^{i\hat\sigma_z\theta_p/2}&\Bigl[e^{-D(s)/2}
\bmat
1\\ -i
\emat
\nonumber\\
&+i\frac{\ep}{\tilde \Delta}{\rm sign}(s)  e^{D(s)/2}
\bmat
1\\ i
\emat
\Bigr]
\ , \\
w^{(2)}(s) =
e^{i \hat\sigma_z\theta_p/2}&e^{D(s)/2}
\bmat
1\\ i
\emat
\ ,
\end{eqnarray}
where $\ep$ is the energy variable,  $\tDlt^{-1} =2 \int_0^{L/2} e^{-D(s)}ds/v_F$ and
\begin{equation}
D(s) =\frac{2}{v_F}\left|\int_0^{s}\Delta(s')ds'\right| \sim \frac{|s|}{\xi}
\ .
\end{equation}

The full wave function near the left end of the wire
being an eigenfunction of the stationary version of Bogolubov~--~de~Gennes equations \eqref{time-BdG} in the main text
can be written as a combination of the above envelopes
with the corresponding oscillating factors $e^{-i\ep t\pm i\bf k \cdot r}$ for the left and right movers
with certain coefficients $a_k^\pm$ and $b_k^\pm$
\begin{eqnarray}
g({\bf r},t) &= e^{-i\ep t +i k_F s}\left[a_L^+w^{(1)}(s)+b_L^+w^{(2)}(s)\right] \nonumber\\
&+e^{-i\ep t-i k_F s}\left[a_L^-w^{(1)}(-s)+b_L^-w^{(2)}(-s)\right] \ .
\end{eqnarray}
A similar expression can be also written near the right end of the wire
by changing the subscripts $L\to R$ and the angle $\theta_p$ from $0$ to $\pi$, which shifts the origin $x\to x-L$ corresponding to $s(\theta_p=0)>0$ and $s(\theta_p=\pi)<0$.
Matching the wave functions of the left and right movers one can find the equations for the coefficients
\begin{eqnarray}\label{SM:b_k^pm_bR}
\phantom{-}i b_R^\pm &= a_L^\pm e^{-D(L/2)\pm i k_F L}\mp \frac{\ep}{\tDlt} a_R^\pm \ , \\
\label{SM:b_k^pm_bL}
-i b_L^\pm &= a_R^\pm e^{-D(L/2)\mp i k_F L}\mp \frac{\ep}{\tDlt} a_L^\pm  \ .
\end{eqnarray}
Here for simplicity we assume the following symmetry $D(x+L/2)-D(L/2) = D(L/2)-D(x-L/2)$ originated from the assumption of a symmetric order parameter $\Delta(L-s)=\Delta(s)$.
As a result, we get a smooth function describing the solution within the interval $|x|<L/2$
\begin{eqnarray}\label{SM:wave_func}
g(x) = \sum_{\eta=\pm 1}e^{-i\ep t + i \eta k_F x}\Bigl[
a_L^\eta e^{ i \eta k_F L/2}e^{-D(x+L/2)/2}
\bmat 
1\\ -i
\emat 
\nonumber\\
+i a_R^\eta e^{-i \eta k_F L/2}e^{-D(x-L/2)/2}
\bmat 
1\\  i
\emat 
\Bigr]  \ .
\end{eqnarray}
At the ends of the wire we should put
\begin{eqnarray}\label{SM:Scatt_matching_-}
g(-L/2) = &\sum_{\eta=\pm 1}\left(
a_L^\eta+b_L^\eta \atop
-i a_L^\eta+ i b_L^\eta
\right)e^{-i\ep t+i \eta 0} \ ,\\
\label{SM:Scatt_matching_+}
g(L/2) = i&\sum_{\eta=\pm 1}\left(
a_R^\eta+b_R^\eta \atop
i a_R^\eta- i b_R^\eta
\right)e^{-i\ep t+i \eta 0} \ ,
\end{eqnarray}
where we marked the left (right) movers by the exponents $e^{\pm i 0}$.
One can see that in the vicinity of the wire ends the wave function exhibits a ``jump'' which occurs at the length scale of the coherence length $\xi$ \cite{Vortex_tunneling_Kopnin_2003, Vortex_tunneling_Kopnin_2007}.

\subsection{Scattering problem}
\begin{figure*}[t]
  \center{
  \includegraphics[width=0.8\textwidth]{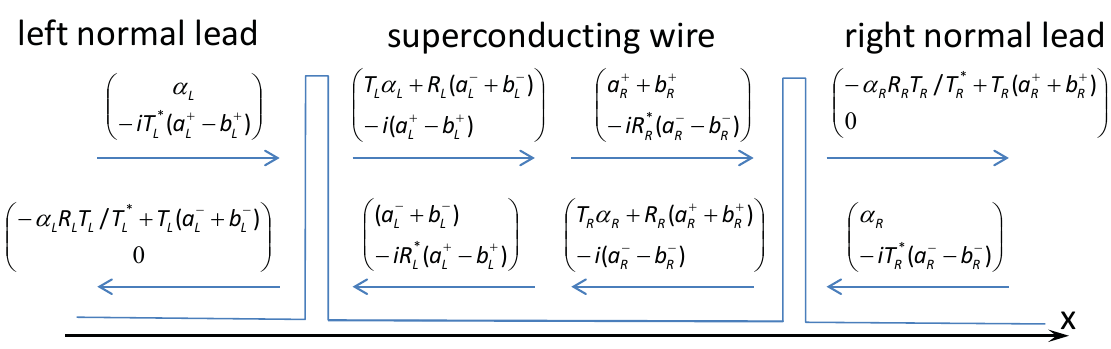}
  \caption{ The scheme shows the electron (upper lines in the brackets) and hole (lower lines) amplitudes in
  the left and right leads and in the vicinity of the interfaces of the superconducting wire in the scattering problem with the amplitudes $\alpha_L$
  ($\alpha_R$) of incoming electronic waves from the left (right) normal lead.
  Right (left) arrows correspond to the factors $e^{_(\pm_)i k_F x}$ in the full wave function \eqref{SM:wave_func}.
  By matching the amplitudes in the latter equation with those shown in this figure one can obtain the matching conditions (\ref{SM:Scatt_matching_-},\ref{SM:Scatt_matching_+}).}
  \label{SM-Fig:Scattering}
  }
  \end{figure*}

As a next step we use
the scattering matrix approach to get the solution of a scattering problem for an electron plane wave $\alpha_{L(R)}e^{\pm i k_F x}$
incident from the left or right normal electrode.
Note that it is enough to consider only incoming electrons, but not holes, if one integrates
over the whole energy interval of the Fermi distribution to calculate the current. Moreover all the sources should be considered
separately by putting only one of them to be non-zero at the same time and summing over all contributions in the observable to avoid any fake interference effects.
Assuming the absence of the electron-hole conversion in the barriers and using the electron-hole symmetry in a superconductor
one can separate complex conjugate electron and hole blocks
in the total scattering matrix of the $k$th barrier $\hat S_k = \left(s_k \quad 0\atop 0 \quad s_k^*\right)$.
We take a standard representation of the unitary matrix $s_k = \left(R_k~\phantom{-R_k^*}T_k\phantom{/T_k^*}\atop T_k~-R_k^* T_k/T_k^* \right)$ which
transforms the incoming electron plane waves from the superconductor (e.g., $a_L^- + b_L^-$ for $k=L$) and from the normal reservoir ($\alpha_L$)
to the outgoing ones ($a_L^+ + b_L^+$ and $u_L=-\alpha_L R_L T_L/T_L^* + T_L (a_L^- + b_L^-)$ for $k=L$) at both interfaces (see Fig.~\ref{SM-Fig:Scattering} for all notations).
Here $R_L = r_L e^{i\phi_L}$, $R_R = r_R e^{-i\phi_R}$ and $T_k$ are the reflection and transmission matrix coefficients, $r_k=\sqrt{1-|T_k|^2}$ and $\phi_k$ are reflection amplitude and phase.

The scattering matrices impose the following boundary conditions on the plane wave amplitudes
\begin{eqnarray}
T_L \alpha_L + R_L (a_L^- + b_L^-) &= a_L^+ + b_L^+ \ , \\
R_L^* (a_L^+ - b_L^+) &= a_L^- - b_L^- \ , \\
T_R \alpha_R + R_R (a_R^+ + b_R^+) &= a_R^- + b_R^-  \ , \\
R_R^* (a_R^- - b_R^-) &= a_R^+ - b_R^+ \ .
\end{eqnarray}

Substituting Eqs.~(\ref{SM:b_k^pm_bR},\ref{SM:b_k^pm_bL}) and introducing the notations $a_k^\pm = e^{\pm i \phi_k/2}(A_k \pm a_k)/2$ one can obtain
the following set of equations
\begin{equation}\label{SM:A_k+a_k_eqs}
\bmat 
\Gamma_L - i\ep   & \omega_0 & 0 & i\tilde\omega_0 \\
-\omega_0 & \Gamma_R - i\ep & i\tilde\omega_0 & 0\\
0 & i\tilde\omega_0 & \frac{\tDlt^2}{\Gamma_L} - i \ep & \omega_0\\
i\tilde\omega_0 & 0 & -\omega_0 & \frac{\tDlt^2}{\Gamma_R} - i\ep\\
\emat 
\bmat 
A_L \\ A_R \\ a_L \\ a_R
\emat 
=
\tDlt
\bmat 
\rho_L\alpha_L \\ \rho_R\alpha_R \\ \frac{1}{\rho_L}\alpha_L \\ \frac{1}{\rho_R}\alpha_R
\emat 
 \ ,
\end{equation}
where $\Gamma_k = \tDlt \rho_k^2$, $\rho_k^2=(1-r_k)/(1+r_k)$, $\omega_0 = \tDlt e^{-D(L/2)}\sin\varphi$,
$\tilde\omega_0 = \tDlt e^{-D(L/2)}\cos\varphi$, $\varphi = k_F L +(\phi_L-\phi_R)/2$.
The phase $\chi_k$ of the transmission coefficients $T_k = |T_k|e^{i\chi_k}$ doesn't affect any measurable quantity,
therefore we choose it equal to $\chi_k=\phi_k/2$ for the sake of simplicity.

A standard recipe to describe the low-frequency ($\omega$) dynamics is to replace the energy $\ep$ by the
time derivative $i\partial/\partial t$. 
In the isolated wire, $\rho_k\to 0$, the fast decaying modes $a_k\sim \alpha_k \rho_k$
disappear as they correspond to the states of the continuous spectrum in the wire and don't satisfy the boundary conditions.
Resulting equations in the closed wire give two energy levels $\ep = \pm \omega_0$ and
correspond to the beating between $A_L$ and $A_R$ in the time domain (see Eqs.~(\ref{SM:time-depend-MBS-eqs_L}, \ref{SM:time-depend-MBS-eqs_R}) below).
Assuming naturally that $\Gamma_k \omega \ll \tDlt^2$ one can find that fast decaying modes
$a_k\approx \alpha_k \rho_k-i\tilde\omega_0 \rho_k^2 A_{k'}\mp \omega_0 \rho_k^2 \rho_{k'} \alpha_{k'}$
corresponding to the continuous spectrum contributions
give small corrections in $e^{-L/\xi}$ to the equations for the low-decaying ones $A_k$. Here and further $k'=R(L)$ for $k=L(R)$.
Indeed, this leads to a relative renormalization of the decay rates $\Gamma_k$ and sources $\rho_k$ by a small values $\sim \omega_0^2/\tDlt^2$ and
to the addition of the $\alpha_{R(L)}$ source proportional to $\tilde \omega_0/\tDlt\sim e^{-L/\xi}$ to the equation for $A_{L(R)}$.
All these terms corresponds to a direct tunneling of electron(s) from the lead to the opposite end of the wire.
Further we neglect these contributions taking into account only a local tunneling from the $k$th leads to the $k$th end of the wire and
considering therefore only first two equations for the amplitudes $A_{L,R}$ of Majorana states in \eqref{SM:A_k+a_k_eqs} without $a_k$.

Transforming the equations to the Schr{\"o}dinger representation one can obtain Eqs.~(\ref{Eq:time-depend-MBS-eqs_L}, \ref{Eq:time-depend-MBS-eqs_R}) from the main text
 \begin{eqnarray}\label{SM:time-depend-MBS-eqs_L}
\left( \frac{\partial}{\partial t} + \Gamma_L \right) A_L +\omega_0 A_R &= F_L e^{-i\ep t}\ ,
 \\
 \label{SM:time-depend-MBS-eqs_R}
\left( \frac{\partial}{\partial t}+\Gamma_R \right) A_R -\omega_0 A_L &= F_R e^{-i\ep t}\ .
\end{eqnarray}
with the choice of sources $F_k = \tDlt \rho_k \alpha_k 
$ appropriate to the replacement $\ep \to i\partial/\partial t$.
Beyond the stationary regime one can consider the parameters $\omega_0$, $\Gamma_k$ and $\rho_k$ to be time-dependent
keeping the Eqs.~(\ref{SM:time-depend-MBS-eqs_L}, \ref{SM:time-depend-MBS-eqs_R}) intact for the typical frequency $\omega$ of the drive small compared to the gap $\omega\ll\tDlt$.

Note that the equations of motion for Majorana amplitudes $A_k$ in the Heisenberg representation (see the first two lines in Eqs.~\eqref{SM:A_k+a_k_eqs})
correspond to the scattering matrix through a scatterer with an internal structure
described in \cite{Scatt_Mat_approach} and applied for the $p$-wave superconducting wire,
e.g., in the Refs.~\cite{MBS_anti-teleport_NilsonAkhBeen2008,Quench_in_MBS_SIS_Been2015}.

\section{Expression for the differential conductance}\label{App:dI_dV}
In this section we consider for simplicity only the case of the non-zero left source $\alpha_L$,
since the results for the right source can be derived using the symmetry $L\leftrightarrow R$.
According to \cite{BTK} the energy resolved contribution to the differential conductance $g_k$ of the $k$th interface can be written
 as a sum of the quasiparticle flows of the left and right moving electrons and holes
with the corresponding signs
\begin{eqnarray}\label{SM:G(E)_def}
g_k(\ep) = 1 - R_k^{e} + R_k^{h} &= |a_k^+ + b_k^+|^2 + |a_k^+ - b_k^+|^2 \nonumber\\
&- |a_k^- + b_k^-|^2 - |a_k^- - b_k^-|^2 \ .
\end{eqnarray}
We used here the conservation of the quasiparticle flow at the interface which results in the unitarity of the scattering matrix.
Substituting the expressions for $b_k^\pm$ (\ref{SM:b_k^pm_bR},\ref{SM:b_k^pm_bL}) and for $a_k^\pm = e^{\pm i \phi_k/2}(A_k \pm a_k)/2$
through the amplitudes $A_k$ and $a_k$ into the Eq.~\eqref{SM:G(E)_def}  one can obtain
\begin{eqnarray}\label{SM:G(E)_general_res}
g_k(\ep) &= \frac{2}{\tDlt^2}{\rm Re}[A_k a_k^*\left(\tDlt^2+\ep^2\right) - A_k A_{k'}^* \ep\tilde\omega_0
-i A_{k'}a_k^* \ep\omega_0 \nonumber\\
&- a_{k'} a_k^* \ep\tilde\omega_0 - i A_k a_{k'}^* \ep\omega_0 + A_{k'} a_{k'}^* (\omega_0^2+\tilde\omega_0^2)
] \ .
\end{eqnarray}
Omitting the terms which are small in the  parameters
$\omega_0/\tDlt,\tilde\omega_0/\tDlt\sim e^{-L/\xi}$ (see the previous section)
one can keep only the first term in the latter equation
\begin{equation}\label{SM:G(E)_res}
g_k(\ep) = 2{\rm Re}[A_k a_k^*] + O\left(A_k^2 \frac{\ep}{\tDlt}e^{-L/\xi}\right) \ .
\end{equation}

In the stationary regime
one can express both $A_k$ and $a_k$ from Eqs.~\eqref{SM:A_k+a_k_eqs}
\begin{eqnarray}
A_k &=& \tDlt\frac{\rho_k\alpha_k(\Gamma_{k'}-i\ep)\mp\omega \rho_{k'}\alpha_{k'}}{(\Gamma_L-i\ep)(\Gamma_R-i\ep)+\omega_0^2}\\
\label{SM:a_k_res}
a_k &\approx& \alpha_k \rho_k 
\end{eqnarray}
and show that
Eq.~\eqref{SM:G(E)_res} transforms into Eq.~\eqref{dc_dI_dV} of the next section, due to incoherence of the left and right sources ($\alpha_L\alpha_R^*\to 0$).
Note that we neglect here all the direct tunneling processes in the wire which give the exponentially small corrections to the Eq.~\eqref{SM:G(E)_res}
in the parameter $L/\xi$.

In general to calculate the zero temperature differential conductance $g_k(eV)$ of the $k$th interface of the wire
in the system with time-dependent parameters one should solve equations (\ref{SM:time-depend-MBS-eqs_L}, \ref{SM:time-depend-MBS-eqs_R}) and substitute
the solutions $A_k$ into the Eq.~\eqref{SM:G(E)_res} together with the expression \eqref{SM:a_k_res} for $a_k$.

\section{Dc differential conductance}\label{App:dc_dI_dV}
Here we consider the dc transport for a constant applied bias $V=V_L-V_R$ using the formalism of the previous section and putting $A_{L,R}\propto e^{-i\ep t}$. As a result we obtain
\begin{equation}\label{dc_dI_dV}
g_{L,R}(\ep) =  \frac{2\Gamma_{L,R} (\Gamma_{R,L} \omega_0^2 +\Gamma_{L,R}(\ep^2+\Gamma_{R,L}^2))}{(\ep^2-\omega_0^2-\Gamma_{L,R}\Gamma_{R,L})^2+\ep^2(\Gamma_{L,R}+\Gamma_{R,L})^2} \ .
\end{equation}
It is convenient to discuss separately the limits $R\to \infty$ and $R\to 0$. For the first limit
 the zero-temperature differential conductance of the device in the symmetric case $\Gamma_L=\Gamma_R$ and $V_R=-V_L$ takes the form
${dI}/{dV}=e^2 g_L(e V/2) /2 \pi$ with $I=I_L=-I_R$ and $g_L(\ep)=g_R(\ep)$.
In the opposite limit of $R\to 0$ similar formulas for the differential conductances hold for each interface separately, i.e.,
${dI_{L,R}}/{dV_{L,R}}=e^2 g_{L,R}(e V_{L,R})/\pi$.
Thus, in both limits we obtain the conductance peak near the zero bias at $eV\sim\omega_0$.
It is this peak which is usually considered { \cite{ZBA_MBS_Mourik2012, ZBA_MBS_Das2012}} as an experimental evidence for the Majorana states in semiconducting wires with
the induced superconducting order.
The nonlocal nature of the Majorana pair reveals itself in the zero bias dip which is of course smeared due to finite rates $\Gamma_{L,R}$
of tunneling to the fermionic baths. As a result, for the exponentially small
splitting $\omega_0$  the dip completely disappears for $\Gamma_{L}\sim\Gamma_{R}\sim\omega_0$
 and  can survive only in a rather exotic limit of strong asymmetry between the couplings to the left and right reservoirs.
The latter situation can be realized, in particular, for the dip in the curve $dI_L/dV_L$ for $R\to 0$ and $\Gamma_R=0$ { \cite{egger,book,Ioselevich_Feigelman2013}}.
A more realistic case with both nonzero tunneling rates and the peak broadening due to the finite temperature and inelastic effects
can make the experimental observation of the $\omega_0$ scale in dc transport difficult.

\end{document}